\documentclass[12pt]{iopart}
\usepackage{iopams}

\usepackage{amssymb}
\usepackage{amsfonts}
\usepackage{bbold}
\usepackage{graphicx}
\usepackage{dcolumn} 
\usepackage{url}
\usepackage{bm}
\usepackage{mathrsfs}
\usepackage[utf8]{inputenc}
\usepackage[colorlinks=true,linkcolor=blue,citecolor=blue,urlcolor=blue]{hyperref}
\usepackage{enumerate}
\usepackage{amsthm}
\usepackage{verbatim}
\usepackage{cite}
\usepackage{bbm}
\usepackage{stmaryrd}
\usepackage{slashed}
\usepackage{revsymb} 
\usepackage{siunitx}
\usepackage{ragged2e}

\newcommand{\DFA}{\address{$^1$Departamento de Física e Astronomia, Faculdade de Ciências da Universidade do Porto, Rua do Campo Alegre s/n, 4169-007, Porto, Portugal.}}
\newcommand{\CFP}{\address{$^3$Centro de Física das Universidades do Minho e do Porto, Rua do Campo Alegre s/n, 4169-007, Porto, Portugal.}}
\newcommand{\UEL}{\address{$^2$Departamento de Física, Universidade Estadual de Londrina, CEP 86051-990, Londrina-PR, Brazil.}}

\begin{document}

\title{Accretion of Generalized Chaplygin Gas onto Cosmologically Coupled Black Holes}
\author{Luís Felipe Reis$^{1,*}$, Mario C. Baldiotti$^2$ and Orfeu Bertolami$^{1,3}$}

\DFA
\UEL
\CFP

\ead{up202209156@up.pt$^*$}
\vspace{10pt}

\date{December 2025}

\begin{abstract}
We study the accretion of cosmic dark fluids, responsible for driving the accelerated expansion of the universe, onto cosmologically coupled black holes. More specifically, we focus on the accretion of the Generalized Chaplygin Gas (GCG). To incorporate the global features of the GCG into this analysis, we employ the McVittie metric, which describes a black hole embedded in an expanding cosmological background. Within this framework, accretion is studied while consistently accounting for the backreaction on the metric components. Using a perturbative approach, we derive an expression for the effective black hole mass and for the evolution of both the black hole and cosmological apparent horizons under accretion. The analysis is performed in two distinct cosmological regimes: first, a matter-dominated era, and subsequently, a de Sitter era. In both cases, it is possible to determine analytically the instant in which accretion begins. For the matter-dominated era, the analytical expression shows that the greater the amount of matter available for accretion, the longer the accretion takes to start.
\end{abstract}

\maketitle

\section{Introduction}

The discovery of the universe's accelerated expansion, initially revealed
through Type Ia supernovae luminosity-distance measurements in the late
1990s \cite{Hamuy:1996ss,Schmidt_1998,Peebles_2003}, has since been corroborated by a multitude of
independent cosmological probes \cite{DESI2018,abbott2022dark, DESI2024}.
Observations of the cosmic microwave background (CMB), baryon acoustic
oscillations (BAO), large-scale structure surveys, and weak gravitational
lensing are broadly consistent with the standard cosmological model ($\Lambda $%
CDM), although a growing consensus about its shortcomings has emerged in recent years \cite{CosmoVerseNetwork:2025alb}.

While the cosmological constant remains the simplest and most
widely adopted model for dark energy, a variety of dynamical models have
been proposed to address the theoretical and observational challenges. Among these, the Generalized Chaplygin Gas (GCG) has attracted attention due to its ability to interpolate between dust-like matter at early times and a dark energy-dominated regime at late times \cite{kamenshchik2001alternative, Bilic:2001cg, bento2002generalized}. 

The GCG model not only provides a compelling observational framework to account for the universe's accelerated expansion and to address the dark matter problem \cite{bento2003wmap, aurich2018concordance, barreiro2008wmap}, but also offers promising avenues for addressing fundamental issues such as the cosmological constant problem \cite{bertolami2023seeding}, the Hubble tension \cite{yang2019dawn}, the Swampland conjectures \cite{bertolami2024sitter}, and inflation through a model inspired by the GCG \cite{Bertolami:2006zg}. 

Despite significant progress in constraining dark energy models, its interaction with compact astrophysical objects remains a relatively unexplored frontier. Black holes, with their intense gravitational fields and central role in galactic evolution,
represent unique laboratories for probing such interactions under extreme conditions. 

Some studies have already explored dark energy and the GCG in the context of compact objects, such as in Refs. \cite{Mazur:2001fv,bertolami2005chaplygin, gorini2009more}. Other works have also investigated  dark energy and the GCG in accretion scenarios, such as in Refs. \cite{babichev2005accretion,rodrigues2012accretion}, although there remains substantial room for further development in this area.

Classical accretion theory has been instrumental in describing the infall of baryonic matter and the associated electromagnetic signatures \cite{bondi1952spherically, petterson1980variations, michel1972accretion, porth2017black}. However, in the context of a universe dominated by a dark energy, it becomes imperative to consider whether and how black holes may accrete this non-standard component.

The accretion of dark energy, particularly when modeled as a perfect fluid
with negative pressure, introduces novel theoretical possibilities. In
scenarios involving phantom energy, that satisfies the condition $p+\rho <-1$, it has been shown that
the accretion process could lead to a decrease in black hole mass, a
counterintuitive result that challenges conventional views of black hole
growth and thermodynamics \cite{babichev2004black}. However,
despite the consistency of such approaches, certain aspects are often
neglected, largely due to the considerable complexity involved in developing
a comprehensive accretion model.

One such critical aspect is the backreaction induced by the accretion process itself. In the standard test-fluid approximation, where the accreting material is assumed too small to significantly modify the black hole’s spacetime and the geometry is treated as fixed, the accreting matter or energy is treated as having negligible influence on the background spacetime geometry. 

Backreaction effects have been explored in steady accretion models where the self-gravity of the accreting fluid modifies the background geometry; for example, the fully relativistic treatment in \cite{malec1994optical} examines how the gravitational influence of matter affects accretion compared to the Bondi limit, and Ref. \cite{Mach:2022wrk} studies steady critical accretion effects on the characteristics of the sonic points.

However, for accretion processes that persist over cosmological timescales or involve significant energy fluxes, the assumption of a fixed black hole metric may no longer be valid. Backreaction effects can, in principle, modify the black hole's horizon, influence the spacetime curvature in the near-horizon region, and potentially alter the global accretion dynamics. To this end, we employ a perturbative approach that enables to find solutions for the evolution of the black hole's apparent horizon.

Another fundamental aspect is that, since accretion is inherently a local phenomenon, while dark energy models are employed to describe a global feature such as the accelerated expansion of the universe, it is essential to adopt a spacetime metric that accommodates both aspects. Specifically, we require a solution that locally describes a spherically symmetric object, while simultaneously incorporating the effects of cosmic expansion. In other words, we seek to study a black hole whose spacetime is not asymptotically flat, but rather asymptotically approaches a flat Friedmann-Lemaitre-Robertson-Walker (FLRW) cosmology.

For this purpose, we adopt the McVittie metric \cite{mcvittie1933mass}. This
spacetime solution has been the subject of ongoing debate for nearly eighty
years regarding its physical consistency in describing black holes embedded
in an expanding universe. However, recent studies have demonstrated that the
McVittie metric not only provides a consistent framework for modeling black
holes that are asymptotically FLRW \cite{kaloper2010mcvittie, lake2011more, nandra2012effect, nandra2012effect1, faraoni2015cosmological}, but also serves as an important foundational model for the
development of more sophisticated and realistic metrics to this end \cite{gao2011black}.

The present work does not aim to delve into the broader debate concerning
the validity and limitations of the McVittie metric. Nevertheless, since
this metric will be employed in the next section to model both the accretion
process and the associated backreaction, the results obtained may offer
valuable insights and contribute to future efforts toward constructing more
complete metrics capable of describing astrophysical objects within an
expanding universe.

This paper is organized as follows: In Section \ref{section2} we analyze
the McVittie metric, with an emphasis on the effective mass associated with
this geometry and on the time-dependent solutions of the apparent horizons.
In Section \ref{section3}, we use a perturbative analysis that has already been developed
that accounts for backreaction effects to study accretion within the McVittie spacetime. In Section \ref{applications}, we apply the results developed in Section \ref{section3} to the GCG, focusing on two distinct analyses: one corresponding to a matter-dominated universe and the other to a de Sitter universe. Finally, in Section \ref{section4}, we present our conclusions and
discuss the implications of the results.

Throughout this work we use units in which $c=G=1$ and adopt the metric signature to be $-+++$.
Also, we will use the notation $^{\cdot }=\partial /\partial t$ and $%
^{\prime }=\partial /\partial r$.

\section{The McVittie metric and the GCG}\label{section2} 
The McVittie metric can be expressed in several equivalent forms; however,
in this work, we adopt the formulation presented in Ref. \cite{kaloper2010mcvittie}.
The line element is given by 
 
\begin{equation}
ds^{2}=-\left( 1-\frac{2M_{0}}{r}-H_{0}^{2}\left( t\right) r^{2}\right)
dt^{2}-\frac{2rH_{0}\left( t\right) }{\sqrt{1-\frac{2M_{0}}{r}}}dtdr 
+\frac{dr^{2}}{1-\frac{2M_{0}}{r}}+r^{2}d\Omega _{2}^{2},  \label{MV metric}
\end{equation}

where $M_{0}$ is the mass parameter associated with the central black hole, $H_{0}\left( t\right)\equiv \dot{a}(t)/a(t) $ is the Hubble parameter, $a(t) $ is the scale factor of the universe, and $d\Omega_{2}^{2}=d\theta ^{2}+\sin ^{2}\theta d\varphi ^{2}$ is the metric element on the $2$-sphere. It is straightforward to verify that, in the limit $M_{0}\rightarrow 0$ and $r\rightarrow \infty $, the metric in Eq. (\ref{MV metric}) reduces to the FLRW metric\footnote{In this work, we consider the McVittie metric in the context of a spatially flat universe.}, while for $H_{0}(t)\rightarrow 0$, it reduces to the Schwarzschild solution. Additionally, it is important to note that when $H_{0}(t)$ is a constant, the metric (\ref{MV metric}) corresponds to the Schwarzschild--de Sitter (SdS) solution.

Since we are not yet accounting for accretion, Eq. (\ref{MV metric})
will be treated as a background metric, denoted by $g_{\mu \nu }^{\left(
0\right) }$, where the upper index '$(0)$' indicates the zeroth-order
approximation in a perturbative expansion. Namely, the order that excludes
accretion effects.

The Einstein field equations for metric (\ref{MV metric}) yield the following
relations: 
\begin{eqnarray}
\label{EFE}
 8\pi \left( T_{0}^{\text{ }0}\right) ^{\left( 0\right) }&=-3H_{0}^{2}(t),
\label{EFE 1} \\
 8\pi \left(
T_{1}^{\text{ }1}\right) ^{\left( 0\right) }&=-3H_{0}^{2}-\frac{2\dot{H}_{0}}{\sqrt{1-\frac{2M_{0}}{r}}}.  \label{EFE 2}
\end{eqnarray}
If we define $\left( T_{\nu }^{\text{ }\mu }\right) ^{\left( 0\right) }$\
such that $\left( T_{0}^{\text{ }0}\right) ^{\left( 0\right) }=-\rho
^{\left( 0\right) }$ and $\left( T_{1}^{\text{ }1}\right) ^{\left( 0\right)
}=p^{\left( 0\right) }$, where $\rho ^{\left( 0\right) }$ and $p^{\left(
0\right) }$ are the energy density and isotropic pressure, respectively, Eq. (\ref%
{EFE 1}) can be readily recognized as the Friedmann equation. This indicates
that the presence of a black hole does not affect the homogeneity of the
energy density distribution. However, Eq. (\ref{EFE 2}) reveals that the
same is not true for the pressure, which acquires a spatial dependence due
to the black hole's influence.

To further elucidate the non-homogeneous nature of the pressure, we consider
the energy conservation equation derived from the Einstein field equations.
This relation is given by 

\begin{equation}
\dot{\rho}^{\left( 0\right) }+3H_{0}\left( p^{\left( 0\right) }+\rho
^{\left( 0\right) }\right) \sqrt{1-\frac{2M_{0}}{r}}=0.  \label{eq conserv}
\end{equation}%
Since both $\rho ^{\left( 0\right) }$ and $H_{0}$ are functions of time
only, we can infer that the term multiplying $3H_{0}$ in Eq. (\ref{eq
conserv}) must also be purely time-dependent. Therefore, we can write: 
\begin{equation}
\left( p^{\left( 0\right) }+\rho ^{\left( 0\right) }\right) \sqrt{1-\frac{%
2M_{0}}{r}}=f\left( t\right) ,  \label{g(t)}
\end{equation}%
where $f\left( t\right) $ is an arbitrary function of time. The authors in Refs.
\cite{nandra2012effect, nandra2012effect1} have shown that for a
matter-dominated universe, it is appropriate to take $f\left( t\right) =\rho
^{\left( 0\right) }$, such that the pressure can be written as%
\begin{equation}
p^{\left( 0\right) }=\frac{\rho ^{\left( 0\right) }}{\sqrt{1-\frac{2M_{0}}{r}%
}}-\rho ^{\left( 0\right) }.  \label{p de rho}
\end{equation}%
In the limit $r\rightarrow \infty $, the pressure in Eq. (\ref{p de rho})
asymptotically approaches the form characteristic of a dust-like fluid.

We can also analyze Eq. (\ref{g(t)}) for a universe governed by the GCG. This fluid is described by an
exotic equation of state of the form  
\begin{equation}
p_{GCG}\left( t\right) =-\frac{A}{\rho _{GCG}^{\alpha }\left( t\right) },
\label{Chap EoS}
\end{equation}%
where $A>0$ and $0<\alpha \leq 1$. The Chaplygin gas corresponds to the choice $\alpha=1$ \cite{kamenshchik2001alternative, bento2002generalized, bento2003wmap}. In the framework of FLRW cosmology, the relativistic
energy conservation associated to the equation of state (\ref{Chap EoS}) leads to an evolving energy density: 
\begin{equation}
\rho _{GCG}\left( t\right) =\left( A+\frac{B}{a^{3\left( 1+\alpha \right)
}\left( t\right) }\right) ^{1/(1+\alpha) },  \label{Chap energy}
\end{equation}%
where $B$ is an integration constant.

Therefore, in the study of a universe governed by the equation of state of the GCG, one
is justified to choose in Eq. (\ref{g(t)}) $f\left( t\right) =\rho ^{\left( 0\right) }-A/\left(
\rho ^{\left( 0\right) }\right) ^{\alpha }$. With this identification, the
pressure can be expressed as: 
\begin{equation}
p^{\left( 0\right) }\left( r,t\right) =-\rho ^{\left( 0\right) }+\frac{1}{%
\sqrt{1-\frac{2M_{0}}{r}}}\left( \rho ^{\left( 0\right) }-\frac{A}{\left(
\rho ^{\left( 0\right) }\right) ^{\alpha }}\right) ,  \label{inhomo pressure}
\end{equation}%
where far from the black hole, the pressure reduces to the standard GCG
equation of state given in Eq. (\ref{Chap EoS}). Consequently, when the
pressure is written in the form of Eq. (\ref{inhomo pressure}), the solution
to the energy conservation equation (\ref{eq conserv}) recovers the same
expression for $\rho ^{\left( 0\right) }$ as given in Eq. (\ref{Chap energy}%
). However, for the sake of generality, we shall work with Einstein's equations (%
\ref{EFE}) in terms of the energy-momentum tensors components $\left( T_{\nu }^{\text{ }%
\mu }\right) ^{\left( 0\right) }$.

Other key aspects of the McVittie metric relevant to our analysis are the
effective mass and the apparent horizon. The effective mass of the black
hole can be computed using the Misner-Sharp mass function $M_{MS}^{\left(
0\right) }$ \cite{misner1964relativistic}, defined by%
\begin{equation}
1-\frac{2M_{MS}^{\left( 0\right) }}{r}=\left( g^{\mu \nu }\right) ^{\left(
0\right) }\partial _{\mu }r\partial _{\nu }r,  \label{Misner Sharp}
\end{equation}%
which, using Eqs. (\ref{MV metric}) and (\ref{EFE 1}), results in 
\begin{equation}
M_{MS}^{\left( 0\right) }\left( t,r\right) =M_{0}+\frac{H_{0}^{2}r^{3}}{2}.
\label{MS function}
\end{equation}

Since the McVittie spacetime is dynamical according to Eq. (\ref{MS function}), a global event horizon cannot, in general, be identified without knowledge of the full future development of the spacetime. It is therefore more appropriate to characterize black hole boundaries quasi-locally in terms of apparent horizons.

In a given spacelike hypersurface, an apparent horizon is defined as the outermost marginally outer trapped surface (MOTS). One may equivalently characterize the apparent horizon in terms of the optical scalars—namely, the expansion $\Theta$, the twist $\omega$, and the shear $\sigma$. In this framework, the apparent horizon is specified by the conditions $\Theta_{+}=0$, $\Theta_{-}<0$, $\omega=0$, and $\sigma=0$, where $\Theta_{+}=0$ corresponds to the vanishing expansion of the outgoing null congruence, while $\Theta_{-}<0$ ensures that the ingoing null congruence remains converging \cite{malec1994optical}.

For a spherically symmetric spacetime given by the metric (\ref{MV metric}), the symmetry two-spheres have area $\mathcal{S} = 4\pi r^{2}$ and, in the coordinates used here, the outgoing and ingoing expansions take the form

\begin{equation}
\Theta_{\pm}=\frac{2}{r} \frac{dr}{ds_{\pm}} \propto \sqrt{1- \frac{2M_{0}}{r}} \left(H_{0}r \pm \sqrt{1- \frac{2M_{0}}{r}}  \right)  \label{optical}.
\end{equation}%
Therefore, the marginally trapped condition $\Theta_{+}=0$ yields
\begin{equation}
H_{0}^{2}\left( r_{h}^{\left( 0\right) }\right) ^{3}-r_{h}^{\left( 0\right)
}+2M_{0}=0,  \label{cubic equation}
\end{equation}%
where $r_{h}^{\left(0 \right)}$ represents the geometrical locus of the apparent horizon.

As shown in Refs. \cite{nickalls1993new, faraoni2012making} the solutions
of the cubic equation above can be expressed as:

\begin{eqnarray}
r_{1}\left( t\right) &=\frac{2}{\sqrt{3}H_{0}\left(
t\right) }\sin \left( \frac{\arcsin \left( 3\sqrt{3}M_{0}H_{0}\left(
t\right) \right) }{3}\right) ,  \label{r1} \\
r_{2}\left( t\right) &=\frac{1}{H_{0}\left( t\right) }%
\cos \left( \frac{\arcsin \left( 3\sqrt{3}M_{0}H_{0}\left( t\right) \right)
}{3}\right)\\ \nonumber 
&-\frac{1}{\sqrt{3}H_{0}\left( t\right) }\sin \left( \frac{\arcsin \left( 3%
\sqrt{3}M_{0}H_{0}\left( t\right) \right) }{3}\right) ,  \label{r2} \\
r_{3}\left( t\right) &=-\frac{1}{H_{0}\left( t\right) }%
\cos \left( \frac{\arcsin \left( 3\sqrt{3}M_{0}H_{0}\left( t\right) \right)
}{3}\right)  \\ \nonumber
&-\frac{1}{\sqrt{3}H_{0}\left( t\right) }\sin \left( \frac{\arcsin \left( 3%
\sqrt{3}M_{0}H_{0}\left( t\right) \right) }{3}\right) .  \label{r3}
\end{eqnarray}

As discussed in Ref. \cite{faraoni2012making}, the solution $r_{3}$ can be
discarded as it is unphysical, since it is always negative. The solution $%
r_{1}$ can be interpreted as the apparent horizon of the black hole, as it
reduces to the Schwarzschild radius $r_{1}\rightarrow 2M_{0}$ in the limit $%
H_{0}\rightarrow 0$. Similarly, $r_{2}$ can be interpreted as the
cosmological horizon, as it approaches $r_{2}\rightarrow 1/H_{0}$ when $%
M_{0}\rightarrow 0$. Henceforth we will denote $%
r_{1}$ by $r_{b}$ and $r_{2}$ by $r_{c}$.

Furthermore, the solutions given by Eqs. (\ref{r1})-(\ref{r2}) impose the
following conditions: for $3\sqrt{3}M_{0}H_{0}<1$, the horizons $r_{b}$ and 
$r_{c}$ are distinct; when $3\sqrt{3}M_{0}H_{0}=1$, the black hole horizon coincides with the cosmological horizon; and when $3\sqrt{3}M_{0}H_{0}>1$,
no horizons exist and the black hole becomes a naked singularity.

Finally, it is also important to compute the expansion velocity of the
apparent horizons. Differentiating (\ref{cubic equation}) with respect to
time yields the evolution velocity of the apparent horizon: 
\begin{equation}
 \dot{r}_{i}=\frac{2H_{0}\dot{H}_{0} r_{i}^{3}}{1-3H_{0}^{2} r^{2}_{i}},  
\label{vAH}
\end{equation}
where $i=b,c$.

This construction should be regarded as a simplified model, serving as a
foundational framework for the development of more physically realistic
approaches, particularly those that incorporate accretion processes. In the
following section, we shall build upon the results presented thus far to
construct a more refined model that accounts not only for accretion but also
for the backreaction effects on the spacetime geometry.

\section{Perturbative approach to backreaction}\label{section3}

The process of accretion within the framework of the McVittie metric has
been investigated in previous works \cite{gao2011black, gao2008does, guariento2012realistic}. The approach
adopted in these studies primarily consists in promoting the mass parameter $%
M_{0}$ in Eq. (\ref{MV metric}) to a time-dependent function.

In the present work, we assume that the accretion of matter induces backreaction effects on the geometry. To investigate this backreaction, we employ a perturbative method as proposed in Ref. \cite{babichev2012backreaction}. We do it so by assuming that accretion induces both a temporal and radial dependence of the mass, as well as an inhomogeneity in the Hubble parameter. Consequently, the metric (\ref{MV metric}) becomes: 

\begin{eqnarray}
ds^{2}&=&-\left( 1-\frac{2M\left( t,r\right) }{r}-H\left( t,r\right)
^{2}r^{2}\right) dt^{2}-\frac{2rH\left( t,r\right) }{\sqrt{1-\frac{2M\left(
t,r\right) }{r}}}dtdr \nonumber \\
&+&\frac{dr^{2}}{1-\frac{2M\left( t,r\right) }{r}}%
+r^{2}d\Omega _{2}^{2}.  \label{MV metric pert}
\end{eqnarray}%

The Einstein equations for the line element (\ref{MV metric pert}) then take the following
form:
\begin{eqnarray}
8\pi \mathcal{T}_{0}^{\text{ }0} &=-3H^{2}-2rHH^{\prime }-\frac{2M^{\prime }}{r^{2}},
\label{EFE 00} \\
8\pi \mathcal{T}_{0}^{\text{ }1} &=2rH\dot{H}+\frac{2\dot{M}}{r^{2}},  \label{EFE 01}
\\
8\pi\mathcal{T}_{1}^{\text{ }0} &=-\frac{2\dot{M}}{\left( r-2M\right) ^{2}},
\label{EFE 10} \\
8\pi \mathcal{T}_{1}^{\text{ }1} &=-\frac{1}{r^{2}\left( r-2M\right) ^{2}}\left[
3r^{2}H^{2}\left( r-2M\right) ^{2}\right. \nonumber\\ &  \left.+2\left( r-2M\right) \left( \left( r-2M\right) M^{\prime }+r^{3}%
\sqrt{1-\frac{2M}{r}}\dot{H}\right) \right.   \nonumber \\
&\left. +2r^{3}H\left( \left( r-2M\right) ^{2}H^{\prime }+\sqrt{1-\frac{2M}{%
r}}\dot{M}\right) \right] .  \label{EFE 11}
\end{eqnarray}
One also finds that the Einstein tensors $G_{2}^{\text{ }2}=G_{3}^{\text{ }%
3}\neq 0$; however, these components yield no independent equations, as they
follow from Eqs. (\ref{EFE 00})-(\ref{EFE 10}).

If we define the function 
\begin{equation}
\mathcal{M}_{MS}\left( t,r\right) =\frac{r}{2}\left( 1-g_{00}\right)
=M\left( t,r\right) +\frac{H\left( t,r\right) ^{2}r^{3}}{2},
\label{MS mass 1}
\end{equation}%
where $g_{\mu \nu }$ is the metric associated with the line element (\ref{MV
metric pert}), then Eqs. (\ref{EFE 00}) and (\ref{EFE 01}) can be rewritten
as 
\begin{equation}
\mathcal{M}_{MS}^{\prime }\left( t,r\right) =\mathcal{B}\left( t,r\right)
\label{M linha}
\end{equation}%
and 
\begin{equation}
\dot{\mathcal{M}}_{MS}\left( t,r\right) =\mathcal{A}\left( t,r\right)
\label{M ponto}
\end{equation}%
where 
\begin{equation}
\mathcal{A}\left( t,r\right) =4\pi r^{2}T_{0}^{\text{ }1}\left( t,r\right) 
\text{ }  \label{A calli}
\end{equation}%
and 
\begin{equation}
\mathcal{B}\left( t,r\right) =-4\pi r^{2}T_{0}^{\text{ }0}\left( t,r\right) .
\label{B calli}
\end{equation}

The system of partial differential equations (\ref{M linha})-(\ref{M ponto})
admits the general solution 
\begin{eqnarray}
\mathcal{M}_{MS}\left( t,r\right) = M_{i} +\int_{t_{0ac}}^{t}\mathcal{A}\left( \bar{t},r_{0ac}\right) ~d%
\bar{t} +\int_{r_{0ac}}^{r_{b}}\mathcal{B}\left( t,\bar{r}\right) ~d\bar{r},   \label{M solution}
\end{eqnarray}
where $t_{0ac}$ is chosen to be the instant in which accretion starts, $r_{0ac}$ is the apparent horizon at time $t=t_{0ac}$, $r_{b}$ is the geometric locus of the black hole apparent horizon, and $M_{i}$ is defined as $M_{i}=M_{MS}^{\left( 0\right)
}\left( t_{0ac},r_{0ac}\right)+M_{ac}\left( t_{0ac},r_{0ac}\right)$, with $M_{MS}^{\left( 0\right)}\left( t,r\right)$ given by Eq. (\ref{MS function}) and $M_{ac}\left( t_{0ac},r_{0ac}\right)$ being the mass of the accreted fluid at instant $ t_{0ac}$ evaluated at $r_{0ac}$, such that $\mathcal{M}_{MS}\left( t_{0ac},r_{0ac}\right) =M_{i} $.

A solution similar to Eq. (\ref{M solution}) appears in Refs. \cite{babichev2012backreaction, campos2025dynamical}. However, in the present context, we are
dealing with a slightly more general scenario, since both the accretion term
and the energy density are allowed to depend on time. Moreover, it is worth
highlighting that working with the McVittie metric allows us to obtain
solutions analogous to that of Refs. \cite{babichev2012backreaction, campos2025dynamical}, but expressed in a physical time coordinate rather than in an
Eddington--Finkelstein--type time coordinate.

The expression for $\mathcal{M}_{MS}$ reveals the contribution of the
accretion of matter, described by $\mathcal{T}_{\nu }^{\text{ }\mu }$, to the initial
effective mass $M_{i}$. The
first integral in (\ref{M solution}) corresponds to the contribution from
the evolution of the accreting matter flux, evaluated in the region bounded
by $r_{0ac}$ over the interval $t-t_{0ac}$. The second integral represents the
time-dependent contribution from the energy density contained between the
initial apparent horizon before accretion, $r_{0ac}$ , and the apparent
horizon displaced due to accretion, $r_{b}$.

With the solution (\ref{M solution}) at hand, we can now effectively compute the effects of accretion. We only need to evaluate the integrals in Eq. (\ref{M solution}) in terms of the energy–momentum tensors. We shall assume that the tensors $\mathcal{T}_{\nu}^{\ \mu}$ satisfying Einstein’s equations (\ref{EFE 00})-(\ref{EFE 11}) correspond to a perfect fluid, given by
\begin{eqnarray}
\mathcal{T}_{\nu }^{\ \mu } = \left({T}_{\nu }^{\ \mu } \right)^{(0)} +
\left( \rho_{ac} + p_{ac} \right) u_{\nu }u^{\mu } + p_{ac}\delta_{\nu }^{\mu },
\label{perf fluid 2}
\end{eqnarray}
where $\left({T}_{\nu }^{\ \mu } \right)^{(0)}$ represents the background fluid defined by equations (\ref{EFE}), $\rho_{ac}$ and $p_{ac}$ are the energy density and the pressure of the accreting fluid, respectively, with $u^{\mu}$ its $4$-velocity.

We shall also assume that the accreting fluid has no angular motion, so that 
$u^{\mu }=\left( u^{0},vu^{0},0,0\right) $, where $v=dr/dt$. Using the
normalization condition for the four-velocity, $u_{\mu }u^{\mu }=-1$, the
relevant components of the energy--momentum tensor (\ref{perf fluid 2}) are
then given by 
\begin{equation}
\mathcal{T}_{0}^{\text{ }0}=-\frac{3H_{0}^{2}\left( t\right) }{8\pi }-\frac{%
\left( \rho _{ac}+p_{ac}\right) \left( g_{00}+g_{01}v\right) }{%
g_{00}+2g_{01}v+g_{11}v^{2}}+p_{ac}  \label{T00 perturbed}
\end{equation}%
and 
\begin{equation}
\mathcal{T}_{0}^{\text{ }1}=-\frac{\left( \rho _{ac}+p_{ac}\right) \left(
g_{00}v+g_{01}v^{2}\right) }{g_{00}+2g_{01}v+g_{11}v^{2}}.
\label{T10 perturbed}
\end{equation}

However, an evident difficulty immediately arises: the metric components are
expressed in terms of the unknown functions $M\left( t,r\right) $ and $%
H\left( t,r\right) $, which appear in Eq. (\ref{MV metric pert}). To address
this, we shall employ the perturbative approach used in Ref. \cite{babichev2012backreaction}. We
assume that the effects of the accreting fluid are much smaller than those
of the background fluid, so that the energy--momentum tensor of the
accreting fluid can be evaluated in the background metric $g_{\mu \nu
}^{\left( 0\right) }$ given by (\ref{MV metric}). Consequently, the tensor $%
\mathcal{T}_{\nu }^{\text{ }\mu }$ and the solution (\ref{M solution}) will
be considered as a first-order perturbation, namely 
\begin{equation}
\mathcal{T}_{\nu }^{\text{ }\mu }\rightarrow \left( T_{\nu }^{\text{ }\mu
}\right) ^{\left( 1\right) },  \label{T (1)}
\end{equation}%
and 
\begin{equation}
\mathcal{M}_{MS}\rightarrow M_{MS}^{\left( 1\right) }.  \label{MMS}
\end{equation}

For simplicity, we can analyze accretion at the apparent horizon of the
black hole. At the black hole's apparent horizon, $g_{00}^{\left( 0\right)
}=0$ and, consequently, from Eq. (\ref{MV metric}), 
\begin{equation}
1-\frac{2M_{0}}{r_{b}^{\left( 0\right) }}=H_{0}^{2}\left( t\right) \left(
r_{b}^{\left( 0\right) }\right) ^{2}.  \label{horiz eq}
\end{equation}%
Eqs. (\ref{T00 perturbed}) and (\ref{T10 perturbed}) then reduce to: 
\begin{eqnarray}
\left( T_{0}^{\text{ }0}\right) ^{\left( 1\right) }\left( t,r_{b}\right) =-%
\frac{3H_{0}^{2}\left( t\right) }{8\pi }+\frac{\left( \rho
_{ac}+p_{ac}\right) H_{0}^{2}\left( t\right) \left( r_{b}^{\left( 0\right)
}\left( t\right) \right) ^{2}}{v-2H_{0}^{2}\left(
t\right) \left( r_{b}^{\left( 0\right) }\left( t\right) \right) ^{2}} +p_{ac} \label{T00 (1)}
\end{eqnarray}%
and%
\begin{equation}
\left( T_{0}^{\text{ }1}\right) ^{\left( 1\right) }\left( t,r_{b}\right) =%
\frac{\left( \rho _{ac}+p_{ac}\right) H_{0}^{2}\left( t\right) \left(
r_{b}^{\left( 0\right) }\left( t\right) \right) ^{2}v}{%
v-2H_{0}^{2}\left( t\right) \left( r_{b}^{\left( 0\right)
}\left( t\right) \right) ^{2}}.  \label{T10 (1)}
\end{equation}

Under this perturbative approximation, Eq. (\ref{M solution})
expressed in terms of the Eqs. (\ref{T00 (1)}) and (\ref{T10 (1)}) is
exact at the initial time $t=t_{0ac}$, such that 
\begin{equation}
g_{00}\left( t_{0ac},r_{0ac}\right) =g_{00}^{\left( 0\right) }\left(
t_{0ac},r_{b}^{\left( 0\right) }\left( t_{0ac}\right) \right) ,
\end{equation}%
and provides an approximation for $t>t_{0ac}$.

We may then analyze the integrals appearing in Eq. (\ref{M solution})
using the definitions from Eqs. (\ref{A calli}) and (\ref{B calli}). We are going to define the first integral in Eq. (\ref{M solution}), after substituting Eq. (\ref{T10 (1)}), as
\begin{equation}
 F\left( t\right) \equiv \int_{t_{0ac}}^{t}\mathcal{A}\left( \bar{t},r_{0ac}\right) ~d\bar{t}.  \label{F(t)}
\end{equation}%

Since our goal is to study the accretion of the GCG, before evaluating the
second integral in Eq. (\ref{M solution}), we anticipate that $\rho _{ac}$
and $p_{ac}$ with which we will work depend only on time. Evidently, the same assumption holds for the
radial velocity of the fluid, $v$, once our goal is to evaluate the
accretion at a specific point, the black hole horizon, i.e., $v(t,r)=v(t,r^{(1)}_{b})$. Consequently, $\left( T_{0}^{\text{ }0}\right) ^{\left( 1\right) }$
given by Eq. (\ref{T00 (1)}), will depend solely on time, and therefore Eq.
(\ref{B calli}) becomes 
\begin{eqnarray}
\int_{r_{0ac}}^{r_{b}^{\left( 1\right) }}\mathcal{B}\left( t,\bar{r}\right) ~d\bar{%
r}=-4\pi \left( T_{0}^{\text{ }0}\left( t\right) \right) ^{\left( 1\right)
}\int_{r_{0ac}}^{r_{b}^{\left( 1\right) }}\bar{r}^{2}~d\bar{r}=\frac{H_{1}^{2}\left( t\right) }{2}\left( r_{b}^{\left( 1\right) }\right) ^{3}+G\left(
t\right), 
\\ \label{int B}
\end{eqnarray}
where we define 
\begin{equation}
H_{1}^{2}\left( t\right) \equiv -\frac{8\pi \left( T_{0}^{\text{ }0}\left(
t\right) \right) ^{\left( 1\right) }}{3}  \label{H1}
\end{equation}%
and 
\begin{equation}
G\left( t\right) \equiv -\frac{H_{1}^{2}\left( t\right) }{2}r_{0ac}^{3}.
\label{G(t)}
\end{equation}

Finally, the solution given in Eq. (\ref{M solution}) can be expressed as 
\begin{equation}
M_{MS}^{\left( 1\right) }\left( t,r\right) =M_{1}\left( t\right) +\frac{%
H_{1}^{2}\left( t\right) }{2}\left( r_{b}^{(1)}\right)^{3},  \label{pert MS mass}
\end{equation}%
where we define 
\begin{equation}
M_{1}\left( t\right) \equiv M_{i}+F\left( t\right)
+G\left( t\right) .  \label{M1(t)}
\end{equation}%
Furthermore, this allows us to compute the evolution of the apparent horizon
displaced due to accretion, $r^{\left( 1\right) }$, through the equation $%
g_{00}^{\left( 1\right) }=0$. Recalling from Eq. (\ref{MS mass 1}) that 
\begin{equation}
 r_{b}^{(1)} =2M_{MS}^{\left( 1\right) }\left(
t,r_{b}^{(1)}\right) ,  \label{rb1}
\end{equation}%
it follows from Eq. (\ref{pert MS mass}) that 
\begin{equation}
H_{1}^{2}\left( t\right) \left( r_{b}^{\left( 1\right) }\right) ^{3}-r_{b}^{\left(
1\right) }+2M_{1}\left( t\right) =0.  \label{eq cube pert}
\end{equation}

This equation is structurally analogous to the cubic Eq. (%
\ref{cubic equation}). Its physically meaningful solutions, in agreement with (\ref{r1}) and (\ref{r2}), are given by 

\begin{eqnarray}
r_{b}^{\left( 1\right) }\left( t\right) &=\frac{2}{\sqrt{3}H_{1}\left(
t\right) }\sin \left( \frac{\arcsin \left( 3\sqrt{3}M_{1}\left( t\right)
H_{1}\left( t\right) \right) }{3}\right) , \nonumber \\  \label{r1b} \\
r_{c}^{\left( 1\right) }\left( t\right) &=\frac{1}{H_{1}\left( t\right) }%
\cos \left( \frac{\arcsin \left( 3\sqrt{3}M_{1}\left( t\right) H_{1}\left(
t\right) \right) }{3}\right)  \nonumber \\
&-\frac{1}{\sqrt{3}H_{1}\left( t\right) }\sin \left( \frac{\arcsin \left( 3%
\sqrt{3}M_{1}\left( t\right) H_{1}\left( t\right) \right) }{3}\right).  \label{r1c}  
\end{eqnarray}
Therefore, within the adopted perturbative framework, we obtain analytical
expressions for the temporal evolution, expressed in terms of a physical
time coordinate, of both the black hole and cosmological apparent horizons
as a result of the accretion process.

Moreover, the conditions required for the solutions in Eqs. (\ref{r1b})-(\ref{r1c}) to be real are similar to those discussed in the previous section for Eqs.
(\ref{r1}) and (\ref{r2}), now for $M_{0}\rightarrow M_{1}\left( t\right) $
and $H_{0}\left( t\right) \rightarrow H_{1}\left( t\right) $.

Naturally, the analytical character of the solutions $r_{b}^{\left( 1\right)
}\left( t\right) $ and $r_{c}^{\left( 1\right) }\left( t\right) $ depends on
whether $M_{1}\left( t\right) $ and $H_{1}\left( t\right) $ can also be
treated analytically. Therefore, we shall investigate the
explicit solutions for specific $\rho _{ac}$ and $p_{ac}$.

\bigskip

\section{Applications} \label{applications}

\subsection{Dark energy accretion}

Before analyzing the GCG, let us first consider a simpler case: the
accretion of a fluid whose equation of state corresponds to that of dark
energy, namely, 
\begin{equation}
p_{ac}=-\rho _{ac}\Rightarrow p_{\Lambda }=-\rho _{\Lambda }.
\end{equation}%
From Eqs. (\ref{F(t)}), (\ref{H1}), and (\ref{M1(t)}), we get 
\begin{eqnarray}
F\left( t\right)  &=0, \\
H_{1}^{2}\left( t\right)  &=H_{0}^{2}+\frac{8\pi \rho _{\Lambda }}{3}, \\
M_{1}\left( t\right)  &=M_{i}-\left( \frac{H_{0}^{2}}{2}%
+\frac{4\pi \rho _{\Lambda }}{3}\right) r_{0ac}^{3}.
\end{eqnarray}%
Therefore, from Eq. (\ref{pert MS mass}), we have%
\begin{equation}
M_{MS}^{\left( 1\right) }\left( t,r_{b}^{\left( 1\right) }\right)
=M_{i}+\left( \frac{H_{0}^{2}}{2}+\frac{4\pi \rho
_{\Lambda }}{3}\right) \left( \left( r_{b}^{\left( 1\right) }\right)
^{3}-r_{0ac}^{3}\right) .  \label{Mass DE}
\end{equation}

A similar equation appears in Ref. \cite{babichev2012backreaction}, but in the case where $H_{0}=0$.
In Ref. \cite{babichev2012backreaction} treatment, Einstein's equations are evaluated in a vacuum
background, that is, $G_{\mu \gamma }\left[ g_{\mu \gamma }^{(0)}\right] =0$%
. In contrast, our zeroth-order solution to Einstein's equations is given by
the McVittie equations.

\bigskip

\subsection{Generalized Chaplygin Gas}

\bigskip

Let us now analyze the accretion of the GCG, whose equation of state is
given by Eq. (\ref{Chap EoS}). From Eq. (\ref{Chap energy}), we identify
two distinct evolutionary phases of the universe. For small time scales, the
universe is matter-dominated such that 
\begin{equation}
\rho^{(matter)} _{GCG}\approx \frac{B^{1/\left( 1+\alpha \right) }}{a^{3}},
\label{GCG matter}
\end{equation}%
while for large time scales, the universe behaves as a de Sitter spacetime,
where 
\begin{equation}
\rho ^{(dS)}_{GCG}\approx A^{1/\left( 1+\alpha \right) }\Rightarrow p^{(dS)}_{GCG}=-\rho^{(dS)}
_{GCG}.  \label{GCG DE}
\end{equation}

Furthermore, there exists an effective equation of state in the intermediate
regime between the matter-dominated and de Sitter phases, which can be
obtained by expanding Eq. (\ref{Chap energy}) as in Ref. \cite{bento2002generalized}:

\begin{eqnarray}
\rho _{GCG}&\approx A^{1/\left( 1+\alpha \right) }+\left( \frac{1}{1+\alpha }%
\right) \frac{B}{A^{\alpha /\left( 1+\alpha \right) }}a^{-3\left( 1+\alpha
\right) }+...,  \label{GCG inter rho} \\
p_{GCG} &\approx-A^{1/\left( 1+\alpha \right) }+\left( 
\frac{\alpha }{1+\alpha }\right) \frac{B}{A^{\alpha /\left( 1+\alpha \right)
    }}a^{-3\left( 1+\alpha \right) }+...   \label{GCG inter p}
\end{eqnarray}

Eqs. (\ref{GCG inter rho})-(\ref{GCG inter p}) are essential, since within
the perturbative framework we are using, the contribution of the accreting
fluid must be much smaller than those of the background fluid.

However, since we are assuming that the GCG drives the cosmic
expansion, thus determining the background metric, while we are also
interested in analyzing its accretion, we shall adopt the following
strategy: we associate the dominant term in Eq. (\ref{GCG inter rho})-(\ref{GCG inter p})
with the background fluid characterized by $H_{0}$, and the remaining term
with the perturbation.

For example, in a de Sitter universe: 
\begin{equation}
a\rightarrow \infty \Rightarrow \left\{ 
\begin{array}{l}
H_{0}=\sqrt{\frac{8\pi \rho^{(dS)}_{GCG}}{3}}=\sqrt{\frac{8\pi A^{1/\left( 1+\alpha \right) }}{3}}, \\ 
\rho _{ac}=\left( \frac{1}{1+\alpha }\right) \frac{B}{A^{\alpha /\left(
1+\alpha \right) }}a^{-3\left( 1+\alpha \right) }, \\ 
p_{ac}=\left( \frac{\alpha }{1+\alpha }\right) \frac{B}{A^{\alpha /\left(
1+\alpha \right) }}a^{-3\left( 1+\alpha \right) }=\alpha \rho_{ac} .%
\end{array}%
\right.  \label{de sitter}
\end{equation}%
For a matter-dominated universe:%
\begin{equation}
a^{(matter)}\Rightarrow \left\{ 
\begin{array}{l}
H_{0}=\sqrt{\frac{8\pi \rho^{(matter)}_{GCG}}{3}}=\frac{2}{3t}, \\ 
\rho _{ac}=A^{1/\left( 1+\alpha \right) }, \\ 
p_{ac}=-A^{1/\left( 1+\alpha \right) }=-\rho _{ac}.%
\end{array}%
\right. .  \label{matter}
\end{equation}

We shall now analyze each case separately.

\bigskip

\subsubsection{Matter-dominated era}

\bigskip

In a universe governed by the GCG during its matter-dominated phase, from Eq. (%
\ref{GCG matter}) we have 
\begin{equation}
H_{0}\left( t\right) =\frac{2}{3t}.  \label{Hubble matter}
\end{equation}%
According to our perturbative construction, we shall assume that the
accreting fluid obeys the subdominant term of the equation of state given in
Eq. (\ref{GCG inter rho})-(\ref{GCG inter p}). Thus, its equation of state reads%
\begin{equation}
\rho _{ac}=-p_{ac}=A^{1/\left( 1+\alpha \right) }. \label{EoS matter}
\end{equation}%
Consequently, from Eqs. (\ref{F(t)}), (\ref{H1}), and (\ref{M1(t)}), we obtain 
\begin{eqnarray}
F\left( t\right)  &=0,  \label{F matter} \\
H_{1}^{2}\left( t\right)  &=\frac{4}{9t^{2}}+\frac{8\pi A^{1/\left(
1+\alpha \right) }}{3},  \label{H1 matter} \\
M_{1}\left( t\right)  &=M_{0}+M_{ac}+\left( H_{0}^{2}\left( t_{0ac}\right)
-H_{1}^{2}\left( t\right) \right) \frac{r_{0ac}^{3}}{2},  \label{M1 matter}
\end{eqnarray}
where, for an energy density given by (\ref{EoS matter}), $M_{ac}=4\pi r_{ac}^{3}\rho_{ac}/3$. Therefore, the effective mass expressed as a first-order perturbation as given by Eq. (\ref%
{pert MS mass}), becomes 
\begin{equation}
M_{MS}^{\left( 1\right) }=M_{i}+\left( \frac{2}{9t^{2}}+%
\frac{4\pi A^{1/(1+\alpha)}}{3}\right) \left( \left( r_{b}^{\left( 1\right)
}\right) ^{3}-r_{0ac}^{3}\right) .  \label{MS1 matter}
\end{equation}

We now study how accretion affects the evolution of the cosmological and
black hole apparent horizons in comparison with the non-accreting case. To
this end, we compare Eqs. (\ref{r1}) and (\ref{r2}) with Eqs. (\ref{r1b})-(\ref{r1c}).

From Eqs. (\ref{r1}) and (\ref{r2}), as discussed in section \ref{section2}, when the equation
\begin{equation}
3\sqrt{3}M_{0}H_{0}\left( t\right) =1  \label{cond 1}
\end{equation}%
is satisfied, this corresponds to the instant at which the cosmological and black hole horizons
form, being equal at that moment. For%
\begin{equation}
3\sqrt{3}M_{0}H_{0}\left( t\right) <1,  \label{cond 2}
\end{equation}%
the horizons become distinct.

Hence, in a matter-dominated universe where the Hubble parameter is given by Eq.
(\ref{Hubble matter}), the horizons form at the instant 
\begin{equation}
t_{0}=2\sqrt{3}M_{0}  \label{t0}
\end{equation}%
and become distinct for $t>t_{0}$. 

Analogously, we can identify the instant $t_{0ac}$ corresponding to the
formation of the horizons due to accretion. Similarly to the non-accreting case, from
Eqs. (\ref{r1b})-(\ref{r1c}) it follows that the horizons are
formed at instant $t_{0ac}$ satisfying    
\begin{equation}
3\sqrt{3}M_{1}\left( t_{0ac}\right) H_{1}\left( t_{0ac}\right) =1.
\label{cond accre}
\end{equation}%
Using Eqs. (\ref{H1 matter}) and (\ref{M1 matter}), Eq. (\ref{cond accre}) admits, as its positive solution, 
\begin{equation}
t_{0ac}=\frac{t_{0}}{\sqrt{1-72\pi M_{0}^{2}A^{1/\left( 1+\alpha
\right) }}}=\frac{t_{0}}{\sqrt{1-72\pi M_{0}^{2} \rho_{ac}}},  \label{t0ac matter}
\end{equation}%
which, in turn, imposes the bound $A<\left( 72\pi M_{0}^{2}\right) ^{-\left(
1+\alpha \right) }$.

It is straightforward to notice that, for $A=0$, that is, when $\rho_{ac}=0$, $t_{0ac}=t_{0}$. However, it is important to emphasize that the delay in the formation time of the apparent horizons is not due to the accretion process itself, but rather to the fact that there is now more matter available for accretion. More precisely, this refers to matter with mass $M_{ac}=4\pi r_{0ac}^{3}\rho_{ac}/3$. 

\begin{figure}[ht]
    \centering
    \includegraphics[width=1 \columnwidth]{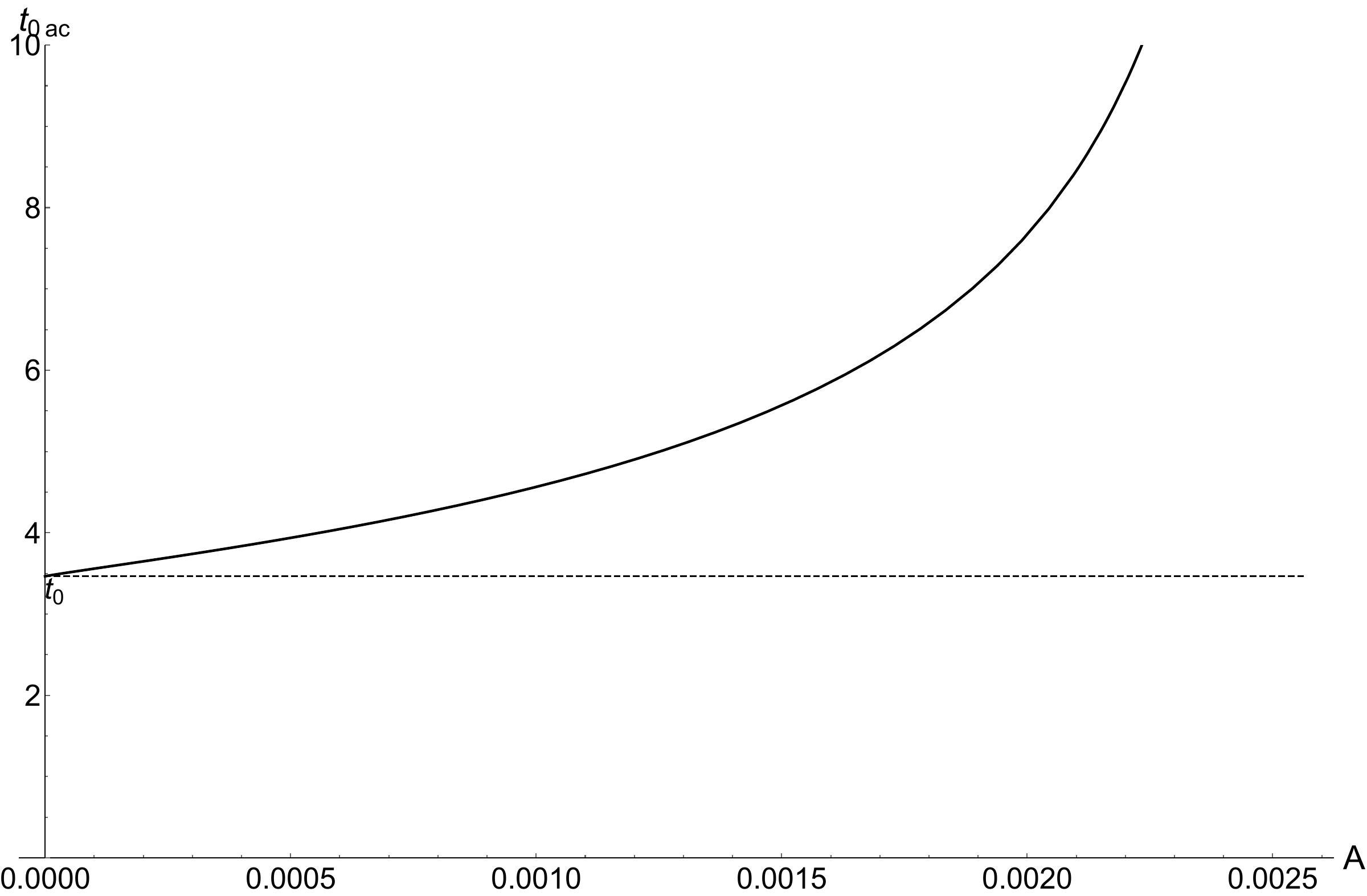}

\caption{\justifying{Apparent black hole and cosmological horizon formation time as a function of $A$ (solid lines), as given by (\ref{t0ac matter}). The dashed line represent the instant of formation of the apparent horizons without available matter for accretion, $t_{0}$, given by \ref{t0}. The mass parameter $M_{0}$ is set to $1$. We also consider $\alpha=0.1$.}}
\label{Figura2}
\end{figure}

From this, we can clearly observe the effect of available matter for accretion on the instant of horizon formation by plotting Eq. (\ref{t0ac matter}), as shown in Fig. \ref{Figura2}. The plot shows that, as $A$ increases, the black hole and cosmological horizons takes longer to form.

This result appears counterintuitive at first. For instance, in static black holes such as the Schwarzschild solution, a greater amount of matter typically leads to faster gravitational collapse and earlier horizon formation. To clarify this result, we now examine the evolution of the horizons.

\begin{figure}[ht]
    \centering
    \includegraphics[width=1\columnwidth]{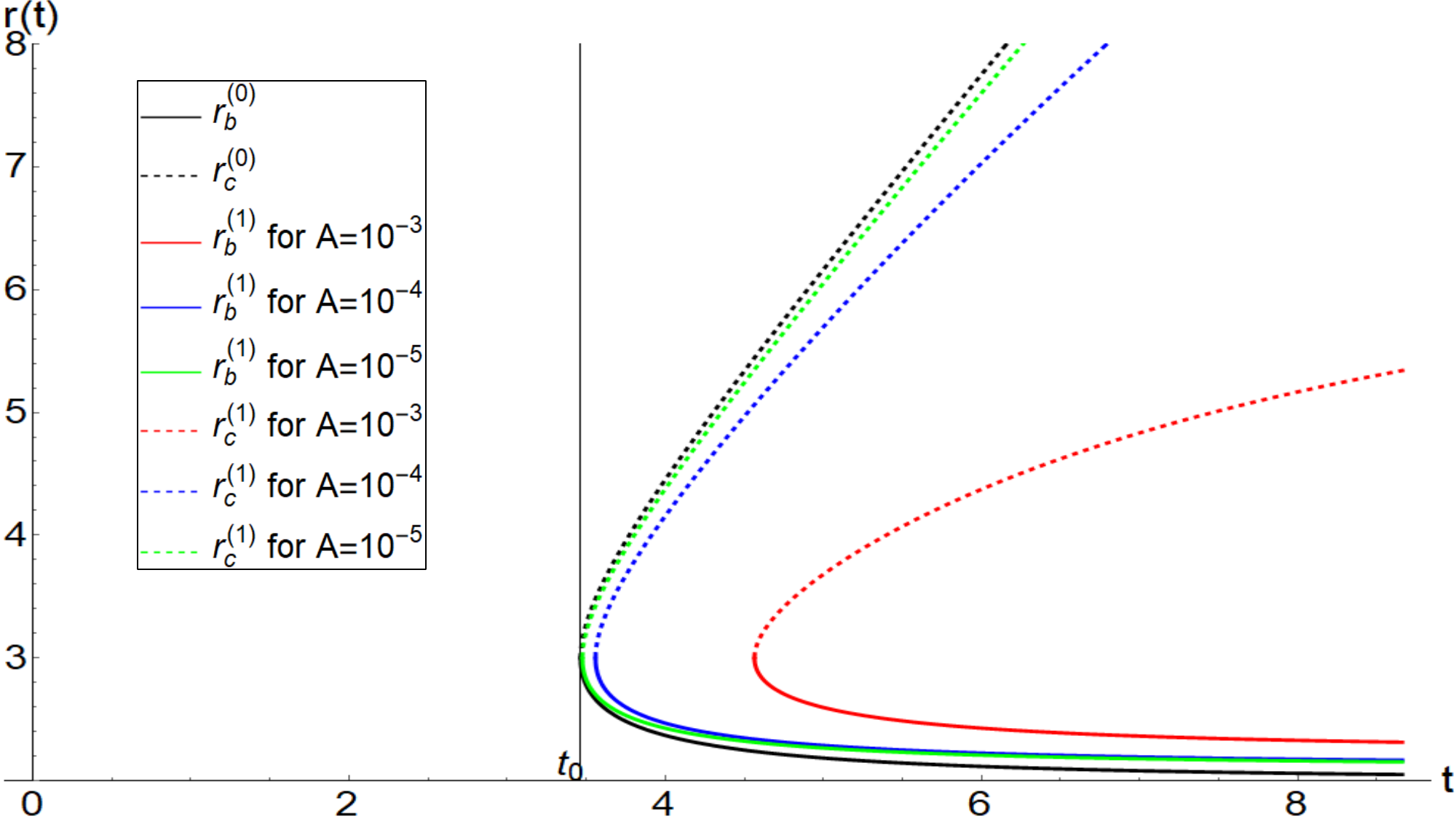}

\caption{\justifying{Temporal evolution of the apparent black hole horizon without available matter for accretion (solid black line) and with available matter for accretion (colored solid lines); and of the cosmological horizons without available matter for accretion (black dashed line) and with available matter for accretion (colored dashed lines) for different values of $A$. The mass parameter $M_{0}$ is set to $M_{0}=1$, such that $t$ is measured in units of $M_{0}$. Also, we consider $\alpha=0.1$.The horizontal dashed line represents the asymptote $r=2M_{0}$. The vertical solid line represents the instant $t_{0}$ at which the horizons form in the case without accretion, as given by Eq. (\ref{t0}).}}

\label{Figura1}
\end{figure}

The horizons obtained from Eqs. (\ref{r1b})-(\ref{r1c}), together with Eqs. (\ref{H1 matter}) and (\ref{M1 matter}), are shown in Fig. \ref{Figura1} for different values of $A$. As illustrated, larger values of $A$, and therefore higher energy densities of matter available to the accretion process, cause the black hole horizon to expand, starting from an initial radius of $3M_{0}$ and settling at constant values greater than the Schwarzschild radius $2M_{0}$ after a sufficiently long time. In contrast, the cosmological horizon decreases with increasing $A$, eventually reaching a steady monotonic regime.

Let us now examine the horizons in their asymptotic stage, beginning with the case without accretion. After a sufficiently long time, such that $t>>t_{0ac}$, the horizons given by Eqs. (\ref{r1}) and (\ref{r2}) behave as    
\begin{eqnarray}
r_{b}^{\left( 0\right) }\left( t\right) &&\approx 2M_{0}\\
r_{c}^{\left( 0\right) }\left( t\right) &&\approx \frac{1}{H_{0}}-M_{0}\approx\frac{3t}{2}-M_{0}. 
\end{eqnarray}
Similarly, for Eqs. (\ref{r1b})-(\ref{r1c}), we find that for time
scales such that $t>>t_{0ac}$, the behavior of the horizons in the presence of
accretion is given by%

\begin{eqnarray}
r_{b}^{\left( 1\right) }\left( t\right) &&\approx 2M_{1}\left( t\right) \approx3M_{0}-36\pi A^{1/(1+\alpha)}M_{0}^{3}, \label{approx r1}  \\ 
r_{c}^{\left( 1\right) }\left( t\right) &&\approx \frac{1}{H_{1}\left(
t\right) }-M_{1}\left( t\right) \approx \left(\frac{4}{9t^{2}}+\frac{4\pi A^{1/(1+\alpha)}}{3} \right)^{-1/2}-\frac{3M_{0}}{2}\nonumber \\ 
&&+18\pi A^{1/(1+\alpha)}M_{0}^{3} \label{approx r2},
\end{eqnarray}
where we used the fact that $r_{0ac} \equiv r_{b}^{(1)}(t_{0ac})=3M_{0}$.

For the black hole horizon with accretion to be larger than in the case without accretion, the condition $A<\left( 12\pi M_{0}^{2}\right) ^{-\left(
1+\alpha \right) }$ must be satisfied, according to Eq. (\ref{approx r1}). However, due to the constraint previously imposed by Eq. (\ref{t0ac matter}), this requirement is automatically fulfilled.

Fig. \ref{Figura1} also shows that, for values of $A$ on the order of $10^{-3}{\si{\meter}}^{-2(1+\alpha)}$, the evolution of the horizons differs significantly from the case without accretion. Therefore, we assume that our perturbative scheme remains valid for an energy density of the accreting fluid of the order of $10^{-4}{\si{\meter}}^{-2(1+\alpha)}$ when compared to that of the background fluid. Therefore, given Eq. (\ref{matter}), this implies that our perturbative approach remains valid only for timescales shorter than 
\begin{equation}
    t=10^{2}\sqrt{\frac{32\pi}{27\rho_{ac}}}. \label{time}
\end{equation}

Furthermore, it is worth noting that the values of the Chaplygin parameter $A$ explored here are chosen for illustrative clarity and exceed those expected from cosmological considerations. Observationally consistent models typically give $10^{-60}{\si{\meter}}^{-2(1+\alpha)} < A < 10^{-54}{\si{\meter}}^{-2(1+\alpha)}$ for $\alpha\leq0.1$, as in Refs. \cite{aurich2018concordance,barreiro2008wmap}. Using enhanced values in our simulations makes the accretion effects numerically visible while preserving the qualitative behavior of the horizon evolution. The physically motivated range of $A$ would simply place these effects at correspondingly smaller amplitudes, without modifying the trends highlighted by our results.

Therefore, although the apparent horizon of the black hole attains larger values due to accretion while the cosmological apparent horizon decreases, as expected, we obtained the counterintuitive result given by Eq. (\ref{t0ac matter}), namely that the greater the amount of accreting matter, the shorter the time required for the black hole horizon to form.

For a cosmologically coupled black hole like the McVittie
solution, the backreaction of accretion on the metric alters this behavior.
The inflow of accreting matter locally increases the energy density and
consequently reduces the rate at which cosmic expansion decelerates, as
shown in Eqs. (\ref{approx r1}) and (\ref{H1 matter}). The resulting higher
effective expansion rate counteracts the growth of the local gravitational
term, namely, the apparent black hole horizon, as expressed in Eqs. (\ref{approx
r1}) and (\ref{M1 matter}), thereby delaying the instant at which the
apparent horizon condition is satisfied.

\bigskip

\subsubsection{de Sitter era}

\bigskip

In a de Sitter--type universe governed by the GCG, the Hubble parameter is
given by 
\begin{equation}
H_{0}=\sqrt{\frac{8\pi A^{1/(1+\alpha) }}{3}}\equiv H_{A}.  \label{HA}
\end{equation}%
The cosmological scale factor, in turn, takes the form%
\begin{equation}
a\left( t\right) =ke^{H_{A}t},  \label{a}
\end{equation}%
where $k$ is a positive constant.

According to our perturbative approach, since the background is determined
by an equation of state of the form $\rho^{(dS)} _{GCG}=-p^{(dS)}_{GCG}=A^{1/\left(
1+\alpha \right) }$, we assume that the accreting fluid corresponds to the
subdominant term in Eq. (\ref{GCG inter rho})-(\ref{GCG inter p}), namely 
\begin{equation}
p_{ac}=\alpha \rho _{ac}=\left( \frac{\alpha }{1+\alpha }\right) \frac{B}{%
A^{\alpha /(1+\alpha) }}a^{-3\left( 1+\alpha \right) }.  \label{p e rho}
\end{equation}

In this case, the flux $F\left( t\right) $ defined in Eq. (\ref{F(t)}) is not, a
priori, equal to zero. Hence, the radial velocity $v$ must be obtained from
the continuity equation  
\begin{equation}
\left(u^{\mu }\right)^{(0)}\partial _{\mu }\rho _{ac}+\left( \rho _{ac}+p_{ac}\right) \nabla
_{\mu }\left(u^{\mu }\right)^{(0)}=0.  \label{cont}
\end{equation}

It is important to note that, within the adopted approximation, the
continuity equation is evaluated in the background metric, i.e. $u^{\mu
}\rightarrow \left( u^{\mu }\right) ^{\left( 0\right) }$. Furthermore, both $%
\rho _{ac}$ and $p_{ac}$ depend only on time, and the
accretion is evaluated at the black hole horizon, such that $r=r_{b}^{\left(
0\right) }\left( t\right) $ and $g_{00}^{\left( 0\right) }=0$. Additionally,
in a dark energy-dominated era, the Hubble parameter is constant and,
consequently, the horizon radius is also constant, given by $r_{b}^{\left(
0\right) }=2M_{0}$. 

A full analytical solution of the continuity equation in this setting is possible in principle but offers limited physical insight. Instead, we focus on the regime in which the radial velocity satisfies $v>>M_{0}H_{A}$, a limit that is both physically relevant for the systems under consideration and leads to expressions that more transparently capture the underlying dynamics. 

As discussed previously, Refs. \cite{aurich2018concordance,barreiro2008wmap} typically give $10^{-60}{\si{\meter}}^{-2(1+\alpha)} < A < 10^{-54}{\si{\meter}}^{-2(1+\alpha)}$ for $\alpha\leq0.1$. If we conservatively assume an initial accretion velocity $v_{0}=0.01c_{s}$, where $c_{s}=\sqrt{{\partial p}/{\partial \rho}}$ is the sound speed derived from the equation of state (\ref{p e rho}), we obtain $v_{0}=0.01 \alpha^{1/2}$.

Hence, the condition $v>>M_{0}H_{A}$ is satisfied for black holes whose mass $M_{0}$ fulfills $M_{0}<<v_{0}/H_{A}$. For $\alpha=0.1$ and $A=10^{-54} {\si{\meter}}^{-2(1+\alpha)}$, this yields $M_{0}<<3.85\times10^{21}{\si{\meter}}$. In the units adopted here, a supermassive black hole with a mass of one billion solar masses corresponds to $M_{0} \sim 10^{12}{\si{\meter}}$, indicating that the inequality $M_{0}<<v_{0}/H_{A}$ is easily satisfied. 

However, since the full time dependence of $v$ is not known, we assume that this approximation holds only over astrophysical timescales. According to Eq. (\ref{p e rho}), the density of the accreting fluid decreases sharply with cosmological time, indicating that the accretion dynamics at cosmological timescales would be significantly altered. 

Therefore, from Eq. (\ref{F(t)}) we get%
\begin{eqnarray}
F\left( t\right) =64\pi H_{A}^{2}M_{0}^{4}(1+\alpha) \int_{t_{0ac}}^{t}%
\rho _{ac}d\bar{t}=\frac{64\pi H_{A}M_{0}^{4}}{3}\left( \rho _{ac0}-\rho
_{ac}\left(t\right) \right)   \label{F DE}
\end{eqnarray}%
where $\rho _{ac0}$ denotes the density given by Eq. (\ref{p e rho}) at the
instant $t_{0ac}$ when accretion begins. 

For the function $H_{1}\left( t\right) $ defined in Eq. (\ref{H1}), we have%
\begin{equation}
H_{1}\left( t\right) =\left( \frac{8\pi }{3}\right) ^{1/2}\sqrt{%
A^{1/(1+\alpha) }-\alpha \rho _{ac}}.
\end{equation}%
The expression above imposes the constraint 
\begin{equation}
A^{1/(1+\alpha) }>\alpha \rho _{ac},  \label{A rho}
\end{equation}%
as expected within our perturbative framework, since the left-hand side is
associated with the background density, whereas the right-hand side
corresponds to the accreting fluid density.

Solving inequality (\ref{A rho}) for $t$, we obtain  
\begin{equation}
t>\ln \left[ \left( \frac{\alpha }{1+\alpha }\frac{B}{A}k^{-3/\left(
1+\alpha \right) }\right) ^{\frac{1+\alpha }{3H_{A}}}\right] .  \label{t0ac}
\end{equation}%
Thus, we set the instant when accretion begins in this phase of the universe
as%
\begin{equation}
t_{0ac}=\ln \left[ \left( \frac{\alpha }{1+\alpha }\frac{B}{A}k^{-3/\left(
1+\alpha \right) }\right) ^{\frac{1+\alpha }{3H_{A}}}\right] .  \label{t0 DE}
\end{equation}

\begin{figure}[ht]
    \centering
    \includegraphics[width=1\columnwidth]{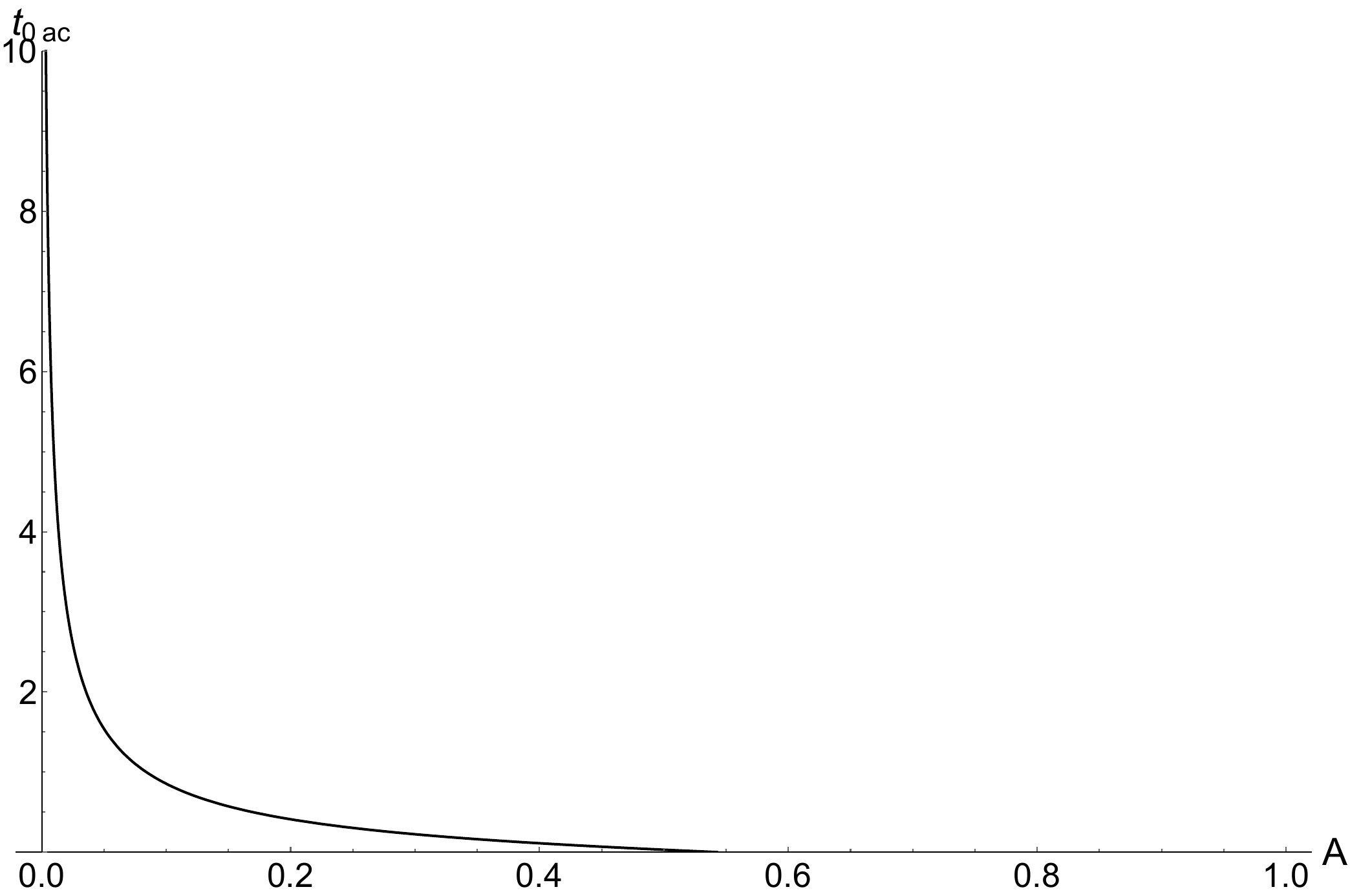}

\caption{\justifying{Apparent black hole and cosmological horizon formation time as a function of $A$, as given by (\ref{t0 DE}). We consider $\alpha=0.1$, $B=1-A$ and $k=0.39$.}}
\label{Figura3}
\end{figure}

This equation allows us to relate the horizon formation time to different
values of $A$. As shown in Fig. \ref{Figura3}, plotting $t_{0ac} $
reveals that, for smaller values of $A$, the horizons take longer to form.

This result is the opposite of that obtained from Eq. (\ref{t0ac matter}).
We can interpret this behavior by recalling that, during the
matter-dominated era, the parameter $A$ is directly associated with the
subdominant term, namely, the energy density of the accreting fluid. In
contrast, in a de Sitter universe, $A$ is directly related to the background
energy density. Consequently, a smaller background energy density implies
less fluid available for accretion, which in turn delays the formation of
the apparent horizons.

Finally, for the function $M_{1}\left( t\right) $ defined in Eq. (\ref{M1(t)}), we
obtain%
\begin{eqnarray}
M_{1}\left( t\right) =M_{0}+\frac{32\pi M^{3}_{0}}{3}\rho_{ac0}+18\pi \alpha M_{0}^{3}\rho _{ac}+\frac{%
64H_{A}M_{0}^{4}\pi }{3}\left( \rho _{ac0}-\rho _{ac}\left( t\right) \right)
, \label{M1 DE}
\end{eqnarray}%
where we used $M_{ac}(t_{0ac})=4\pi r^{3}_{0ac}\rho_{ac0}/3$.

Therefore, since at large times $r_{b}^{\left( 1\right) }\approx
2M_{1}\left( t\right) $ and $r_{c}^{\left( 1\right) }\approx \left(
1/H_{1}\left( t\right) \right) -M_{1}\left( t\right) $, and given that $\rho
_{ac}\left( t\right) $ decreases with time according to Eq. (\ref{p e rho}), the
black hole horizon becomes larger in the presence of accretion than in its
absence, while the cosmological horizon becomes correspondingly smaller.

\section{Conclusions}\label{section4}
In this work, we investigated the accretion of cosmic dark fluids, particularly the GCG, onto cosmologically coupled black holes described by the McVittie metric. By employing a perturbative formalism that takes into account the backreaction of accretion on the metric components, we derived analytical expressions for the effective black hole mass and for the evolution of both the black hole and cosmological apparent horizons. This framework allowed us to analyze how the local accretion process interacts with the global cosmological background in different epochs of the universe.

In the matter-dominated era, our results reveal a somewhat counterintuitive behavior. One might expect that the presence of additional matter available for accretion would accelerate the black hole formation process. However, due to the backreaction effects inherent to the McVittie metric, the situation is reversed. The inflow of matter locally increases the energy density, which in turn reduces the deceleration of the cosmic expansion. As a result, the effective Hubble parameter becomes larger, counteracting the local gravitational collapse. This dynamic causes the apparent horizon of the black hole to form later for higher values of $A$, that is, when the available matter for accretion is greater.

The analysis of Eq. (\ref{t0ac matter}) confirms that the horizon formation time $t_{0ac}$ increases with $A$, which means that the cosmological coupling and the backreaction jointly act to delay the black hole horizon formation compared to the non-accreting case.

In contrast, during the de Sitter phase, the behavior of horizon formation exhibits the opposite dependence on $A$. In this regime, the parameter $A$ is directly tied to the background energy density of the universe rather than to the accreting fluid. Consequently, a smaller value of $A$ corresponds to a lower cosmological energy density, implying less available fluid to be accreted. This reduced inflow slows down the local mass growth, thereby delaying the formation of the horizons. Hence, while in the matter-dominated universe a higher $A$ delays the formation time due to backreaction, in the de Sitter regime, it is a smaller $A$ that produces a similar delay. This complementary behavior highlights the subtle interplay between local and global dynamics in cosmologically coupled systems.

The formalism developed here is sufficiently general to be extended to a variety of related scenarios. In particular, a more detailed investigation of the backreaction effects on the fluid dynamics itself would be of considerable interest, especially concerning the structure of critical sonic points and the transonic nature of the flow in a cosmological background \cite{malec1994optical, Mach:2022wrk, Karkowski:2005zn}. Moreover, the analysis may be generalized to other perfect fluids coupled to the expansion of the universe through alternative equations of state. Such extensions would allow one to assess the robustness of the qualitative features reported here.

Overall, since the McVittie metric provides a bridge between cosmological expansion and local gravitational collapse, our formalism could be particularly relevant for the study of primordial black holes, whose formation and evolution are naturally embedded in an expanding universe. Understanding how accretion and backreaction operate in such early-universe scenarios could shed light on the role of cosmologically coupled black holes in structure formation and in the broader dynamics of the dark sector.

\section*{Acknowledgments}
The work of L.F.R. is supported by FCT (Fundação para a Ciência e Tecnologia, Portugal)
through the grant 2024.06413.BD. The work of O.B. is partially supported by FCT through
the project 2024.00252.CERN. Both authors are partially supported by UID/04650 - Centro
de Física das Universidades do Minho e do Porto. M. C. B. thanks the University of Porto for
hosting part of this study.

\section*{Bibliography}
\bibliographystyle{unsrt.bst}
\bibliography{references}

\begin{thebibliography}{10}

\bibitem{Hamuy:1996ss}
M.~Hamuy, M.~M. Phillips, N.~B. Suntzeff, R.~A. Schommer, J.~Maza, and R.~Aviles.
\newblock {The Hubble diagram of the Calan/Tololo type IA supernovae and the value of H0}.
\newblock {\em Astron. J.}, 112:2398, 1996.

\bibitem{Schmidt_1998}
B.~P. Schmidt et~al.
\newblock {The High Z supernova search: Measuring cosmic deceleration and global curvature of the universe using type Ia supernovae}.
\newblock {\em Astrophys. J.}, 507:46--63, 1998.

\bibitem{Peebles_2003}
P.~J.~E. Peebles and B.~Ratra.
\newblock {The Cosmological Constant and Dark Energy}.
\newblock {\em Rev. Mod. Phys.}, 75:559--606, 2003.

\bibitem{DESI2018}
T.~M.~C. Abbott et~al.
\newblock {Dark Energy Survey year 1 results: Cosmological constraints from galaxy clustering and weak lensing}.
\newblock {\em Phys. Rev. D}, 98(4):043526, 2018.

\bibitem{abbott2022dark}
T.~M.~C. Abbott et~al.
\newblock {Dark Energy Survey Year 3 results: Cosmological constraints from galaxy clustering and weak lensing}.
\newblock {\em Phys. Rev. D}, 105(2):023520, 2022.

\bibitem{DESI2024}
A.~G. Adame et~al.
\newblock {DESI 2024 VI: cosmological constraints from the measurements of baryon acoustic oscillations}.
\newblock {\em JCAP}, 02:021, 2025.

\bibitem{CosmoVerseNetwork:2025alb}
E.~Di~Valentino et~al.
\newblock {The CosmoVerse White Paper: Addressing observational tensions in cosmology with systematics and fundamental physics}.
\newblock {\em Phys. Dark Univ.}, 49:101965, 2025.

\bibitem{kamenshchik2001alternative}
A.~Y. Kamenshchik, U.~Moschella, and V.~Pasquier.
\newblock {An Alternative to quintessence}.
\newblock {\em Phys. Lett. B}, 511:265--268, 2001.

\bibitem{Bilic:2001cg}
N.~Bilic, G.~B. Tupper, and R.~D. Viollier.
\newblock {Unification of dark matter and dark energy: The Inhomogeneous Chaplygin gas}.
\newblock {\em Phys. Lett. B}, 535:17--21, 2002.

\bibitem{bento2002generalized}
M.~C. Bento, O.~Bertolami, and A.~A. Sen.
\newblock {Generalized Chaplygin gas, accelerated expansion and dark energy matter unification}.
\newblock {\em Phys. Rev. D}, 66:043507, 2002.

\bibitem{bento2003wmap}
M.~C. Bento, O.~Bertolami, and A.~A. Sen.
\newblock {WMAP constraints on the generalized Chaplygin gas model}.
\newblock {\em Phys. Lett. B}, 575:172--180, 2003.

\bibitem{aurich2018concordance}
R.~Aurich and S.~Lustig.
\newblock {On the concordance of cosmological data in the case of the generalized Chaplygin gas}.
\newblock {\em Astropart. Phys.}, 97:118--129, 2018.

\bibitem{barreiro2008wmap}
H.~Li, W.~Yang, and Y.~Wu.
\newblock {Constraint on the generalized Chaplygin gas as an unified dark fluid model after Planck 2015}.
\newblock {\em Phys. Dark Univ.}, 22:60--66, 2018.

\bibitem{bertolami2023seeding}
O.~Bertolami.
\newblock {Seeding the vacuum with entropy: the Chaplygin-like vacuum hypothesis}.
\newblock {\em Class. Quant. Grav.}, 40(17):177002, 2023.

\bibitem{yang2019dawn}
W.~Yang, S.~Pan, S.~Vagnozzi, E.~Di~Valentino, D.~F. Mota, and S.~Capozziello.
\newblock {Dawn of the dark: unified dark sectors and the EDGES Cosmic Dawn 21-cm signal}.
\newblock {\em JCAP}, 11:044, 2019.

\bibitem{bertolami2024sitter}
O.~Bertolami, R.~Potting, and P.~M. S{\'a}.
\newblock {The de Sitter Swampland Conjectures in the Context of Chaplygin-Inspired Inflation}.
\newblock {\em Universe}, 10(7):271, 2024.

\bibitem{Bertolami:2006zg}
O.~Bertolami and V.~Duvvuri.
\newblock {Chaplygin inflation}.
\newblock {\em Phys. Lett. B}, 640:121--125, 2006.

\bibitem{Mazur:2001fv}
Pawel~O. Mazur and Emil Mottola.
\newblock {Gravitational Condensate Stars: An Alternative to Black Holes}.
\newblock {\em Universe}, 9(2):88, 2023.

\bibitem{bertolami2005chaplygin}
O.~Bertolami and J.~Paramos.
\newblock {The chaplygin dark star}.
\newblock {\em Phys. Rev. D}, 72:123512, 2005.

\bibitem{gorini2009more}
V.~Gorini, A.~Yu. Kamenshchik, U.~Moschella, O.~F. Piattella, and A.~A. Starobinsky.
\newblock {More about the Tolman-Oppenheimer-Volkoff equations for the generalized Chaplygin gas}.
\newblock {\em Phys. Rev. D}, 80:104038, 2009.

\bibitem{babichev2005accretion}
E.~Babichev, V.~Dokuchaev, and Y.~Eroshenko.
\newblock {The Accretion of dark energy onto a black hole}.
\newblock {\em J. Exp. Theor. Phys.}, 100:528--538, 2005.

\bibitem{rodrigues2012accretion}
M.~G. Rodrigues and A.~E. Bernardini.
\newblock {Accretion of non-minimally coupled generalized Chaplygin gas into black holes}.
\newblock {\em Int. J. Mod. Phys. D}, 21:1250075, 2012.

\bibitem{bondi1952spherically}
H.~Bondi.
\newblock {On spherically symmetrical accretion}.
\newblock {\em Mon. Not. Roy. Astron. Soc.}, 112:195, 1952.

\bibitem{petterson1980variations}
J.~A. Petterson, J.~Silk, and J.~P. Ostriker.
\newblock Variations on a spherically symmetrical accretion flow.
\newblock {\em Mon. Not. Roy. Astron. Soc.}, 191(3):571--579, 1980.

\bibitem{michel1972accretion}
F.~C. Michel.
\newblock {Accretion of matter by condensed objects}.
\newblock {\em Astrophys. Space Sci.}, 15(1):153--160, 1972.

\bibitem{porth2017black}
O.~Porth, H.~Olivares, Y.~Mizuno, Z.~Younsi, L.~Rezzolla, M.~Moscibrodzka, H.~Falcke, and M.~Kramer.
\newblock {The black hole accretion code}.
\newblock {\em Comput. Astrophys. Cosmol.}, 4(1):1, 2017.

\bibitem{babichev2004black}
E.~Babichev, V.~Dokuchaev, and Yu. Eroshenko.
\newblock {Black hole mass decreasing due to phantom energy accretion}.
\newblock {\em Phys. Rev. Lett.}, 93:021102, 2004.

\bibitem{malec1994optical}
E.~Malec and N.O. Murchadha.
\newblock Optical scalars and singularity avoidance in spherical spacetimes.
\newblock {\em Phys. Rev. D}, 50(10):R6033, 1994.

\bibitem{Mach:2022wrk}
P.~Mach and E.~Malec.
\newblock {Steady critical accretion onto black holes: Self-gravity and sonic point characteristics}.
\newblock {\em Phys. Rev. D}, 105(10):104012, 2022.

\bibitem{mcvittie1933mass}
G.~C. McVittie.
\newblock {The mass-particle in an expanding universe}.
\newblock {\em Mon. Not. Roy. Astron. Soc.}, 93:325--339, 1933.

\bibitem{kaloper2010mcvittie}
N.~Kaloper, M.~Kleban, and D.~Martin.
\newblock {McVittie's Legacy: Black Holes in an Expanding Universe}.
\newblock {\em Phys. Rev. D}, 81:104044, 2010.

\bibitem{lake2011more}
K.~Lake and M.~Abdelqader.
\newblock {More on McVittie's Legacy: A Schwarzschild - de Sitter black and white hole embedded in an asymptotically $\Lambda$CDM cosmology}.
\newblock {\em Phys. Rev. D}, 84:044045, 2011.

\bibitem{nandra2012effect}
R.~Nandra, A.~N. Lasenby, and M.~P. Hobson.
\newblock {The effect of a massive object on an expanding universe}.
\newblock {\em Mon. Not. Roy. Astron. Soc.}, 422:2931--2944, 2012.

\bibitem{nandra2012effect1}
R.~Nandra, A.~N. Lasenby, and M.~P. Hobson.
\newblock {The effect of an expanding universe on massive objects}.
\newblock {\em Mon. Not. Roy. Astron. Soc.}, 422:2945--2959, 2012.

\bibitem{faraoni2015cosmological}
V.~Faraoni.
\newblock {\em {Cosmological and Black Hole Apparent Horizons}}, volume 907.
\newblock Springer, 2015.

\bibitem{gao2011black}
C.~Gao, X.~Chen, Y.~Shen, and V.~Faraoni.
\newblock {Black Holes in the Universe: Generalized Lemaitre-Tolman-Bondi Solutions}.
\newblock {\em Phys. Rev. D}, 84:104047, 2011.

\bibitem{misner1964relativistic}
C.~W. Misner and D.~H. Sharp.
\newblock {Relativistic equations for adiabatic, spherically symmetric gravitational collapse}.
\newblock {\em Phys. Rev.}, 136:B571--B576, 1964.

\bibitem{nickalls1993new}
R.~W.~D. Nickalls.
\newblock A new approach to solving the cubic: Cardan’s solution revealed.
\newblock {\em Math. Gazzete}, 77(480):354--359, 1993.

\bibitem{faraoni2012making}
V.~Faraoni, Andres~F. Zambrano~M., and R.~Nandra.
\newblock {Making sense of the bizarre behaviour of horizons in the McVittie spacetime}.
\newblock {\em Phys. Rev. D}, 85:083526, 2012.

\bibitem{gao2008does}
C.~Gao, X.~Chen, V.~Faraoni, and Y.~Shen.
\newblock {Does the mass of a black hole decrease due to the accretion of phantom energy?}
\newblock {\em Phys. Rev. D}, 78:024008, 2008.

\bibitem{guariento2012realistic}
D.~C. Guariento, M.~Fontanini, A.~M. da~Silva, and E.~Abdalla.
\newblock {Realistic fluids as source for dynamically accreting black holes in a cosmological background}.
\newblock {\em Phys. Rev. D}, 86:124020, 2012.

\bibitem{babichev2012backreaction}
E.~Babichev, V.~Dokuchaev, and Yu. Eroshenko.
\newblock {Backreaction of accreting matter onto a black hole in the Eddington-Finkelstein coordinates}.
\newblock {\em Class. Quant. Grav.}, 29:115002, 2012.

\bibitem{campos2025dynamical}
T.~L. Campos, C.~Molina, and M.~C. Baldiotti.
\newblock {Dynamical Black Holes and Accretion-Induced Backreaction}.
\newblock {\em Universe}, 11(7):202, 2025.

\bibitem{Karkowski:2005zn}
J.~Karkowski, B.~Kinasiewicz, P.~Mach, E.~Malec, and Z.~Swierczynski.
\newblock {Universality and backreaction in a general-relativistic accretion of steady fluids}.
\newblock {\em Phys. Rev. D}, 73:021503, 2006.

\end{thebibliography}

\end{document}